\documentclass[12pt]{article}
 \usepackage{graphicx}
 \usepackage[cp1251]{inputenc} %% Windows
 \usepackage[russian]{babel}   %%

\begin{document}
 \noindent {\footnotesize\it Astronomy Letters, 2015, Vol. 41, No. 9, pp. 473--488.}
 \newcommand{\dif}{\textrm{d}}

 \noindent
 \begin{tabular}{llllllllllllllllllllllllllllllllllllllllllllll}
 & & & & & & & & & & & & & & & & & & & & & & & & & & & & & & & & & & & & & \\\hline\hline
 \end{tabular}

  \vskip 1.0cm
  \centerline{\bf Determination of the Galactic Rotation Curve from OB Stars}
  \bigskip
  \centerline{V.V. Bobylev and A. T. Bajkova}
  \bigskip
  \centerline{\small Pulkovo Astronomical Observatory, St. Petersburg,  Russia}
  \bigskip
  \bigskip
{\bf Abstract}—We consider three samples of O- and B-type stars
from the solar neighborhood 0.6--4 kpc for which we have taken the
distances, line-of-sight velocities, and proper motions from
published sources. The first sample contains 120 massive
spectroscopic binaries. O stars with spectroscopic distances from
Patriarchi et al. constitute the second sample. The third sample
consists of 168 OB3 stars whose distances have been determined
from interstellar calcium lines. The angular velocity of Galactic
rotation at the solar distance $\Omega_0,$ its two derivatives
$\Omega'_0$ and $\Omega''_0,$ and the peculiar velocity components
of the Sun $(U,V,W)_\odot$ are shown to be well determined from
all three samples of stars. They are determined with the smallest
errors from the sample of spectroscopic binary stars and the
sample of stars with the calcium distance scale. The fine
structure of the velocity field associated with the influence of
the Galactic spiral density wave clearly manifests itself in the
radial velocities of spectroscopic binary stars and in the sample
of stars with the calcium distance scale.

 %DOI: 10.1134/S1063773715080010

\section*{INTRODUCTION}
Young massive stars of spectral types O and B are visible from
great distances from the Sun. Therefore, they are an important
tool for studying the structure and kinematics of the Galaxy
(Moffat et al. 1998; Zabolotskikh et al. 2002; Avedisova 2005;
Popova and Loktin 2005). Since OB stars do not recede far from
their birthplaces in their lifetime, they trace well the spiral
structure of the Galaxy (Efremov 2011; Coleiro and Chaty 2013;
Vall\'ee 2014; Hou and Han 2014).

So far the distances to active star-forming regions, where OB
stars are concentrated, have been determined by the kinematic
method (Anderson et al. 2012), which has a low accuracy (Mois\'es
2011). The spectroscopic and photometric distance estimation
methods can be used only for a relatively small solar neighborhood
with a radius of 4--5 kpc. However, several factors presently
reduce the accuracy of these methods. First, there are virtually
no direct measurements of the parallaxes for supergiants in the
Hipparcos catalogue (1997) with an accuracy of at least 10\%. Such
measurements are needed to establish a reliable calibration.
Second, the uncertainty in establishing the spectral type and
luminosity class is great for high-luminosity stars. According to
Wegner (2007), the first two factors cause, for example, the
absolute magnitude dispersion for the sequences of Ia or Iab
supergiants on the Hertzsprung–Russell diagram to be about
1.5--2.0 magnitudes (this dispersion is only about 0.3 magnitude
for main-sequence stars). Third, for the application of
photometric methods (for example, based on Str\"omgren
photometry), there are simply no data for distance determinations.

The appearance of infrared observations, in particular, the 2MASS
catalogue (Skrutskie et al. 2006), has contributed to an increase
in the accuracy of these methods when the interstellar extinction
is taken into account. Using the photometric characteristics from
this catalogue, Patriarchi et al. (2003) determined the
spectroscopic distances for 184 O stars. One of our goals is to
study the Galactic kinematics using data on these stars.

Spectroscopic binaries are of interest in their own right. They
usually have a long history of observations. The systemic
line-of-sight velocities, spectral classification, and photometry
are well known for many of them. Having analyzed the detached
spectroscopic binaries with known spectroscopic orbits selected on
condition that the masses and radii of both components were
determined with errors of no more than 3\%, Torres et al. (2010)
showed their spectroscopic distances to agree with the
trigonometric ones within error limits of no more than 10\%. In
Eker et al. (2014), the number of such binaries was increased by
67\%; it is 257. However, the fraction of young massive binaries
among them is small (about 20 binaries).

Previously (Bobylev and Bajkova 2013a), we presented a kinematic
database containing information about 220 massive young stars. The
stars are no farther than 3--4 kpc from the Sun; about 100 of them
are spectroscopic binary and multiple systems whose components are
massive OB stars, while the rest are B stars from the Hipparcos
catalogue with spectral types no later than B2.5 and errors in the
trigonometric parallaxes no more than 10\%. All stars have
estimates of their proper motions, line-of-sight velocities, and
distances, which allows their space velocities to be analyzed. A
significant fraction of the stars from this sample (162 stars)
belongs to the Gould Belt (with heliocentric distances
$r<0.6$~kpc). Previously (Bobylev and Bajkova 2013a, 2014b), we
analyzed the kinematics and spatial distribution of these nearby
stars. In contrast, the number of more distant stars (58 stars at
$r>0.6$~kpc) should be increased for a reliable determination of
the parameters describing the Galactic kinematics.

At present, there are estimates of the distances to OB stars made
by the spectroscopic method from the broadening of interstellar Ca
II, Na I, or K I absorption lines (Megier et al. 2005, 2009), with
this scale having been reconciled with the trigonometric
parallaxes (Megier et al. 2005). Quite recently, new measurements
of stars performed by this method have been published
(Galazutdinov et al. 2015).

The goal of this paper is to expand the kinematic database on
massive young star systems, to compare the various distance
scales, and to determine the parameters of the Galactic rotation
and spiral density wave based on the data obtained.

\section*{METHOD}

From observations we know three projections of the stellar
velocity: the line-of-sight velocity $V_r$ as well as the two
velocities $V_l=4.74r\mu_l\cos b$ and $V_b=4.74r\mu_b$ directed
along the Galactic longitude $l$ and latitude $b$ and expressed in
km s$^{-1}$. Here, the coefficient 4.74 is the ratio of the number
of kilometers in an astronomical unit to the number of seconds in
a tropical year, and $r$ is the heliocentric distance of the star
$r$ in kpc. The proper motion components $\mu_l\cos b$ and $\mu_b$
are expressed in milliarcseconds per year (mas yr$^{-1}$). The
velocities $U,V,$ and $W$ directed along the rectangular Galactic
coordinate axes are calculated via the components $V_r,V_l,$ and
$V_b:$
 \begin{equation}
 \begin{array}{lll}
 U=V_r\cos l\cos b-V_l\sin l-V_b\cos l\sin b,\\
 V=V_r\sin l\cos b+V_l\cos l-V_b\sin l\sin b,\\
 W=V_r\sin b                +V_b\cos b,
 \label{UVW}
 \end{array}
 \end{equation}
where $U$ is directed from the Sun to the Galactic center, $V$ is
in the direction of Galactic rotation, and $W$ is directed toward
the north Galactic pole. We can find two velocities, $V_R$
directed radially away from the Galactic center and $V_{circ}$
orthogonal to it and pointing in the direction of Galactic
rotation, based on the following relations:
 \begin{equation}
 \begin{array}{lll}
  V_{circ}= U\sin \theta+(V_0+V)\cos \theta, \\
       V_R=-U\cos \theta+(V_0+V)\sin \theta,
 \label{VRVT}
 \end{array}
 \end{equation}
where the position angle $\theta$ is calculated as
$\tan\theta=y/(R_0-x),$ while $x$ and $y$ are the rectangular
Galactic coordinates of the star. $\Omega_0$ is the angular
velocity of Galactic rotation at the solar distance $R_0,$ the
parameters $\Omega^{\prime}_0$ and $\Omega^{\prime\prime}$ are the
corresponding derivatives of the angular velocity, and
$V_0=|R_0\Omega_0|.$

The velocities $V_R$ and $W$ are virtually independent of the
pattern of the Galactic rotation curve. However, to analyze any
periodicities in the tangential velocities, it is necessary to
determine a smoothed Galactic rotation curve and to form the
residual velocities $\Delta V_{circ}.$ As experience shows, to
construct a smooth Galactic rotation curve in a wide range of
distances $R,$ it is usually sufficient to know two derivatives of
the angular velocity, $\Omega^\prime_0$ and
$\Omega^{\prime\prime}_0$. Note that all three velocities $V_R,$
$\Delta V_{circ},$ and $W$ must be freed from the peculiar solar
velocity $U_\odot,V_\odot,W_\odot$.

To determine the parameters of the Galactic rotation curve, we use
the equations derived from Bottlinger’s formulas in which the
angular velocity $\Omega$ was expanded in a series to terms of the
second order of smallness in $r/R_0:$
\begin{equation}
 \begin{array}{lll}
 V_r=-U_\odot\cos b\cos l-V_\odot\cos b\sin l\\
 -W_\odot\sin b+R_0(R-R_0)\sin l\cos b\Omega^\prime_0
 +0.5R_0(R-R_0)^2\sin l\cos b\Omega^{\prime\prime}_0,
 \label{EQ-1}
 \end{array}
 \end{equation}
 \begin{equation}
 \begin{array}{lll}
 V_l= U_\odot\sin l-V_\odot\cos l-r\Omega_0\cos b\\
 +(R-R_0)(R_0\cos l-r\cos b)\Omega^\prime_0
 +0.5(R-R_0)^2(R_0\cos l-r\cos b)\Omega^{\prime\prime}_0,
 \label{EQ-2}
 \end{array}
 \end{equation}
 \begin{equation}
 \begin{array}{lll}
 V_b=U_\odot\cos l\sin b + V_\odot\sin l \sin b\\
 -W_\odot\cos b-R_0(R-R_0)\sin l\sin b\Omega^\prime_0
    -0.5R_0(R-R_0)^2\sin l\sin b\Omega^{\prime\prime}_0,
 \label{EQ-3}
 \end{array}
 \end{equation}
where $R$ is the distance from the star to the Galactic rotation
axis,
  \begin{equation}
 R^2=r^2\cos^2 b-2R_0 r\cos b\cos l+R^2_0.
 \end{equation}
The influence of the spiral density wave in the radial, $V_R,$ and
residual tangential, $\Delta V_{circ},$ velocities is periodic
with an amplitude of $\sim$10 km s$^{-1}.$ According to the linear
theory of density waves (Lin and Shu 1964), it is described by the
following relations:
 \begin{equation}
 \begin{array}{lll}
       V_R =-f_R \cos \chi,\\
 \Delta V_{circ}= f_\theta \sin\chi,
 \label{DelVRot}
 \end{array}
 \end{equation}
where
 \begin{equation}
\chi=m[\cot(i)\ln(R/R_0)-\theta]+\chi_\odot
 \end{equation}
is the phase of the spiral density wave ($m$ is the number of
spiral arms, $i$ is the pitch angle of the spiral pattern,
$\chi_\odot$ is the radial phase of the Sun in the spiral density
wave); $f_R$ and $f_\theta$ are the radial and tangential velocity
perturbation amplitudes, which are assumed to be positive.

In the next step, we apply a spectral analysis to study the
periodicities in the velocities $V_R$ and $\Delta V_{circ}$. The
wavelength $\lambda$ (the distance between adjacent spiral arm
segments measured along the radial direction) is calculated from
the relation
\begin{equation}
 \frac{2\pi R_0}{\lambda} = m\cot(i).
 \label{a-04}
\end{equation}
Let there be a series of measured velocities $V_{R_n}$ (these can
be both radial, $V_R$ and residual tangential, $\Delta
V_{\theta},$ velocities), $n=1,\dots,N$, where $N$ is the number
of objects. The objective of our spectral analysis is to extract a
periodicity from the data series in accordance with the adopted
model describing a spiral density wave with parameters
$f_R,\lambda (i)$ and $\chi_\odot$.

Having taken into account the logarithmic character of the spiral
density wave and the position angles of the objects $\theta_n,$
our spectral (periodogram) analysis of the series of velocity
perturbations is reduced to calculating the square of the
amplitude (power spectrum) of the standard Fourier transform
(Bajkova and Bobylev 2012):
\begin{equation}
 \bar{V}_{\lambda_k} = \frac{1} {N}\sum_{n=1}^{N} V^{'}_n(R^{'}_n)
 \exp\Bigl(-j\frac {2\pi R^{'}_n}{\lambda_k}\Bigr),
 \label{29}
\end{equation}
where $\bar{V}_{\lambda_k}$ is the $k$th harmonic of the Fourier
transform with wavelength  $\lambda_k=D/k, D$ is the period of the
series being analyzed,
 \begin{equation}
 \begin{array}{lll}
 R^{'}_{n}=R_{\circ}\ln(R_n/R_{\circ}),\\
 V^{'}_n(R^{'}_n)=V_n(R^{'}_n)\times\exp(jm\theta_n).
 \label{21}
 \end{array}
\end{equation}
The algorithm of searching for periodicities modified to properly
determine not only the wavelength but also the amplitude of the
perturbations is described in detail in Bajkova and Bobylev
(2012).

%%%%%%%%%%%%%%
 \begin{table}[p]                                     % t~1.
 \caption[]{\small Bibliographic information about the stars}
 \begin{center}
  \label{t:01}
 %\small
 \footnotesize\baselineskip=0.1ex
 \begin{tabular}{|r|l|r|l|}\hline
      Star & $V_\gamma$, $~\mu,~$ $dist$&    Star & $V_\gamma$, $~\mu,~$ $dist$ \\\hline

 MWC 1       &  Sb9,  L07, G12    & Herschel 36 & Ar10,  UC4, Ar10  \\
 V745 Cas    & Ch14a, L07, Ch14a  & V411 Ser    &  Sb9,  L07, G12   \\
 BM Cas      &  Sb9,  L07, Pop77  & mu. Sgr     &  Sb9,  L07, BB15  \\
 HD 12323    &  Sb9,  UC4, Gal15  & 16 Sgr      & Ma14,  L07, G12   \\
 V622 Per    & Mal07, UC4, Cap02  &  MY Ser     & Ib13,  L07, Ib13  \\
 V615 Cas    & MJ14, MJ14, MJ14   &  QR Ser     &  S09,  L07, S09   \\
 V482 Cas    &  Sb9, Tyc2, Mc03   &  RY Sct     &  Sb9,  L07, Sm02  \\
 MY Cam      & Lr14, Tyc2, Lr14   &  MCW 715    &  Sb9,  L07, Hut71 \\
 KS Per      &  Sb9, L07,  BB15   &  V337 Aql   & Tu14, Tu14, Tu14  \\
 HD 31617    &  Sb9, L07,  Gui12  &  MWC 314    & Lo13, Tyc2, Car10 \\
 HD 31894    &  Sb9, L07,  Gui12  &  V1765 Cyg  &  Sb9,  L07, G12   \\
 eps Aur     &  Sb9, L07,  Gui12  &  V380 Cyg   &  Sb9,  L07, Tk14  \\
 ADS 4072AB  &  Sb9, L07,  Pat03  &  HD 124314  & SIMB,  L07, G12   \\
 3 Pup       &  Sb9, L07,  RR94   &  V2107 Cyg  &  B14,  L07, B14   \\
 HD 72754    &  Sb9, L07,  L07    &  ALS 10885  &  G06,  L07, Pat03 \\
 ALS 1803    &  Sb9, L07,  Pat03  &  V470 Cyg   &  Sb9,  L07, Com12 \\
 ALS 1834    &  Sb9, Tyc2, Pat03  &  WR 140     &  Sb9, Dz09, Do05  \\
 V560 Car    &  Sb9, Tyc2, R01    &  V404 Cyg   & MJ14, MJ14, MJ14  \\
 V346 Cen    &  Sb9, UC4,  Ka10   &  Schulte 73 & Ki09,  UC4, Ki09  \\
 KX Vel      & Ma14, L07,  Ma14   &  V382 Cyg   &  Sb9,  L07, YY12  \\
 HD 97166    & Ma14, UC4,  Ma14   &  V2186 Cyg  & Ki12,  UC4, Ki12  \\
 HD 115455   & Ma14, UC4,  Ma14   &  MT91 771   & Ki12,  UC4, Ki07  \\
 HD 123590   & Ma14, L07,  Av84   &  VV Cep     &  Sb9,  L07, L07   \\
 HD 150136   &  S13, UC4,  S13    &  V2174 Cyg  &  Sb9,  L07, Bol78 \\
 HD 152234   &  Sb9, L07,  G12    &  ~~~~4U 2206+54 & StZ14, Tyc2, SIMB  \\
 HD 152246   & Na14, L07,  Na14   &  V446 Cep   & Ch14b, L07, Ch14b  \\
 Braes 144   & SIMB, Tyc2, G12    &  MWC 656    & Cas14, L07, Cas14 \\
 V1297 Sco   & SIMB, Tyc2, G12    &  GT Cep     & Ch15,  L07, Ch15  \\
 Pismis 24-1 & MW07, Tyc2, MW07   &  V373 Cas   &  Sb9,  L07, Hl87  \\
 HD 163892   & Ma14, L07,  CaII   &  ALS 12775  &  Sb9,  L07, BB15  \\
 HR 6716     &  Sb9, L07,  St99   &  VZ Cen     & Wil13, L07, BB15  \\ \hline
 \end{tabular}
 \end{center}
 {
  \footnotesize\baselineskip=0.1ex
1. The sources of line-of-sight velocities: Sb9 (Pourbaix et al.
2004); Ch14a, Ch14b, Ch15 (\c{C}hakirli et al. 2014a, 2014b,
2015); Ar10 (Arias et al. 2010); Ma14 (Mayer et al. 2014); G06
(Gontcharov 2006); Mal07 (Malchenko et al. 2007); Ib13 (Ibanoglu
et al. 2013); MJ14 (Miller-Jones 2014); S09, S13 (Sana et al.
2009, 2013); Lr14 (Lorenzo et al. 2014); Tu14 (T\"uys\"uz et al.
2014); StZ14 (Stoyanov et al. 2014); Ki07, Ki09, Ki12 (Kiminki et
al. 2007, 2009, 2012); Lo13 (Lobel et al. 2013); B14 (Baki\c{s} et
al. 2014); Na14 (Nasseri et al. 2014); Cas14 (Casares et al.
2014); MW07 (Maiz-Apell\'aniz et al. 2007); Wil13 (Williams et al.
2013); SIMB (SIMBAD database).

2. The sources of proper motions: L07 (van Leeuwen 2007); UC4
(UCAC4, Zacharias et al. 2012); Tyc2 (Tycho-2, Hog et al. 2000);
Dz09 (Dzib and Rodriguez 2009).

3. The sources of distances: G12 (Gudennavar et al. 2012); CaII
(Megier et al. 2009); Pat03 (Patriarchi et al. 2003); Pop77
(Popper 1977); Gal15 (Galazutdinov et al. 2015); Cap02 (Capilla
and Fabregat 2002); Mc03 (McSwain 2003); Sm02 (Smith et al. 2002);
Av84 (Avedisova and Kondratenko 1984); Bol78 (Bolton and Rogers
1978); Hut71 (Hutchings and Redman 1971); Gui12 (Guinan 2012);
Car10 (Carmona et al. 2010); R01 (Rauw et al. 2001); RR94 (Rovero
and Ringuelet 1994); Com12 (Comer\'on and Pasquali 2012); Do05
(Dougherty et al. 2005); Hl87 (Hill and Fisher 1987); St99
(Stickland and Lloyd 1999); Tk14 (Tkachenko et al. 2014); Ka10
(Kaltcheva and Scorcio 2010); BB15 (our estimates); YY12
(Ya\c{s}arsoy and Yakut 2012). }
      \end{table}
%%%%%%%%%%%%%%%%%%%%%%%%%%%%%%%%%%%%%%%%%%%%%%%%%%%%%%%%%%%%%%%%%

%%%%%%%%%%%%%%%%%%%%%%%%%%%%%%%%%%%%%%%%%%%%%%%%%%%%%%%%%%%%%%%%%%%
 \begin{figure}
 {\begin{center}
 \includegraphics[width=80.0mm]{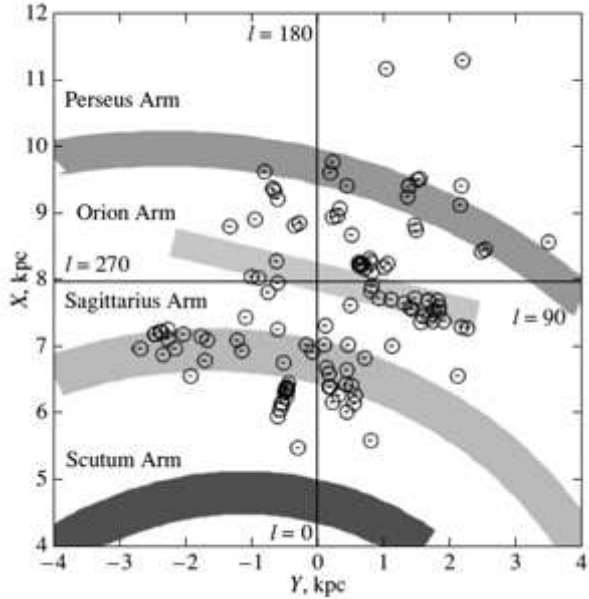}
 \caption{
Distribution of 120 spectroscopic binary stars in projection onto
the Galactic $XY$ plane; the Sun is at the intersection of the
dashed lines; the fragments of the four-armed spiral pattern are
indicated; the position of the local spur, the Orion arm, is
marked. }
  \label{f1}
 \end{center} }
 \end{figure}
%%%%%%%%%%%%%%%%%%%%%%%%%%%%%%%%%%%%%%%%%%%%%%%%%%%%%%%%%%%%%%%%%%%
%%%%%%%%%%%%%%%%%%%%%%%%%%%%%%%%%%%%%%%%%%%%%%%%%%%%%%%%%%%%%%%%%%%
 \begin{figure}
 {\begin{center}
 \includegraphics[width=90.0mm]{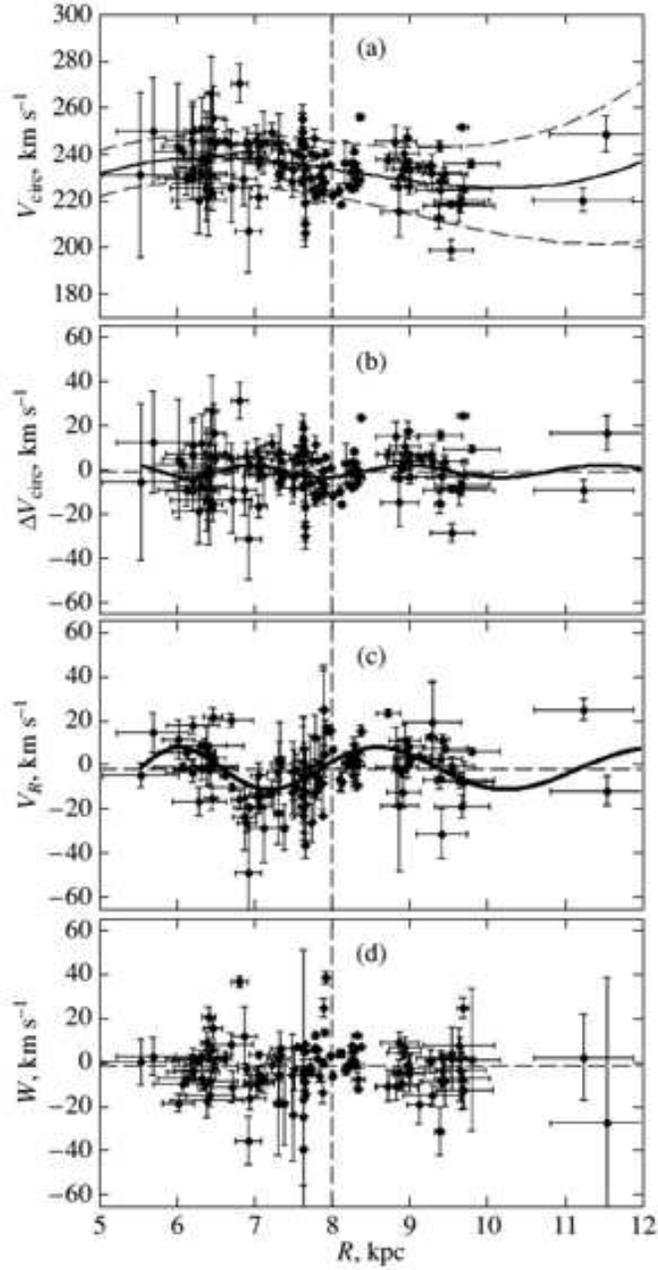}
 \caption{
(a) Galactic rotation curve constructed from the sample of 120
spectroscopic binary stars with parameters (13) with an indication
of the boundaries of the 1$\sigma$ confidence intervals; residual
rotation velocities $\Delta V_{circ}$ (b), radial velocities $V_R$
(c), and vertical velocities $W$ (d) of stars versus $R;$ the
vertical dashed line marks the Sun’s position. }
  \label{f2}
 \end{center} }
 \end{figure}
%%%%%%%%%%%%%%%%%%%%%%%%%%%%%%%%%%%%%%%%%%%%%%%%%%%%%%%%%%%%%%%%%%%
%%%%%%%%%%%%%%%%%%%%%%%%%%%%%%%%%%%%%%%%%%%%%%%%%%%%%%%%%%%%%%%%%%%
 \begin{figure}
 {\begin{center}
 \includegraphics[width=60.0mm]{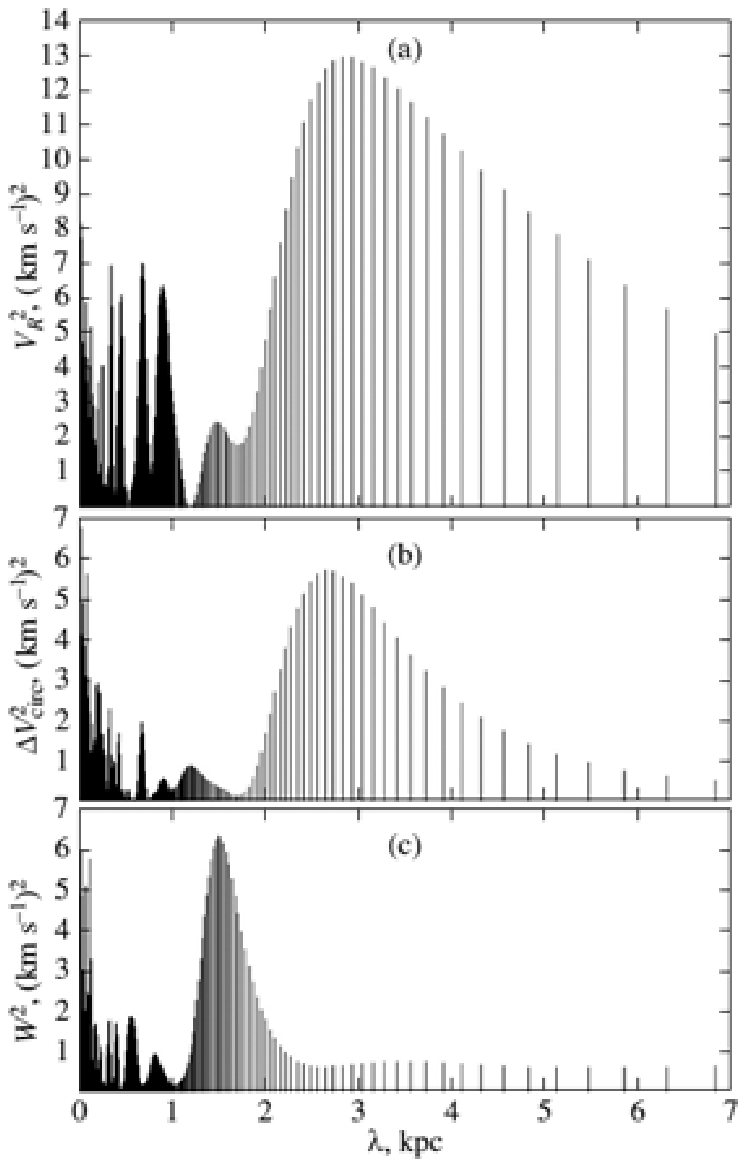} %%Spectr-SB.eps}
 \caption{
Power spectra of the radial, $V_R$, residual tangential, $\Delta
V_{circ},$ and vertical, $W,$ velocities for the sample of
spectroscopic binary stars. }
  \label{f3}
 \end{center} }
 \end{figure}
%%%%%%%%%%%%%%%%%%%%%%%%%%%%%%%%%%%%%%%%%%%%%%%%%%%%%%%%%%%%%%%%%%%

\section*{DATA}\label{Data}
\subsection*{The Sample of Spectroscopic Binary OB Stars}\label{SB}
To select young massive distant spectroscopic binaries, we used
the SB9 database (Pourbaix et al. 2004), which contains references
to the spectroscopic orbit determinations up until 2014. We
considered binaries with spectral types of the primary component
no later than B2.5 and various supergiants with luminosity classes
Ia and Iab. Thus, we are interested in stars with masses of more
than 10 solar masses.

For a preliminary distance estimate, we consulted various sources,
for example, the compiled database by Gudennavar et al. (2012) and
the FUSE (Far Ultraviolet Spectroscopic Explorer) survey (Bowen et
al. 2008). Finally, the publications of various authors that
determine orbits are an important source of information.

We did not consider stars nearer than 0.6 kpc, where the Gould
Belt dominates, and stars farther than 4 kpc, where the proper
motion errors are great. We added 62 more binaries that possess no
properties of ``runaway'' stars (their residual velocities do not
exceed 40--50 km s$^{-1}$) to the 58 already described in Bobylev
and Bajkova (2013a). Bibliographic information about these
binaries is given in the table. There are cases where there is no
orbit in the SB9 catalogue, while it is specified in the SIMBAD
database that the star is a spectroscopic binary, for example,
HD\,124314.

For a number of stars, we calculated the distances by ourselves.
The abbreviation BB15 is used for them in the table. Here, we
calculated the distances using the photometric magnitudes from the
SIMBAD database based on the well-known relation
\begin{equation}
 \log r = 1+0.2((V-M_v)-A_v),
 \label{99}
\end{equation}
where $A_v=3.2 E(B-V),$ $E(B-V)=(B-V)-(B-V)_0.$ We took the
absolute magnitudes of the stars $M_v$ from Wegner (2007), where
they were determined from Hipparcos stars for all spectral types
and luminosity classes, and the unreddened colors of the stars
$(B-V)_0$ from Straizys (1977).

The final sample of spectroscopic binaries contains 120 stars.
Their distribution in the Galactic $XY$ plane is presented in
Fig.~1. The figure shows a fragment of the global four-armed (one
of the four arms, the Outer one, is not visible here) spiral
pattern with a pitch angle of 13$^\circ$. The pattern was
constructed in our previous paper (Bobylev and Bajkova 2014a),
where we analyzed the distribution of Galactic masers with
measured trigonometric parallaxes. The line segment between the
Perseus and Carina--Sagittarius Arms indicates the Local Arm model
(Bobylev and Bajkova 2014b). As can be seen from the figure, the
distribution of stars agrees well with the plotted spiral pattern.
Unfortunately, there are few stars in the third quadrant in the
Perseus Arm. In this paper, we take the Galactocentric distance of
the Sun to be $R_0=8.0\pm0.4$~kpc.

\subsection*{O Stars from the List by Patriarchi et al. (2003)}
Patriarchi et al. (2003) determined the spectroscopic distances
for 184 O-type stars; photometric data from the 2MASS catalogue
were used to take into account the interstellar extinction. This
sample consists mostly of single stars, although there are also
spectroscopic binaries in it.

The proper motions from the Hipparcos catalogue revised by van
Leeuwen (2007) or from Tycho-2 (Hog et al. 2000) and line-of-sight
velocities (according to the SIMBAD database) are known for 156
stars from this list. However, the information is complete only
for 101 of them, i.e., their distances, proper motions, and
line-of-sight velocities are known simultaneously. The
distribution of 156 O stars in projection onto the Galactic $XY$
plane is given in Fig. 4. Just as in Fig. 1, the spiral arms are
plotted. The agreement of the stars from this sample with the
plotted spiral pattern is noticeably poorer than that for the
stars in Fig. 1. Here, we see a larger dispersion of the
coordinates in the Carina--Sagittarius Arm toward the Galactic
center $(l=0^\circ),$ in the Perseus Arm toward the Galactic
anticenter $(l=180^\circ),$ and, finally, the Orion Arm segment is
noticeably stretched in a direction near $l=75^\circ$.

%%%%%%%%%%%%%%%%%%%%%%%%%%%%%%%%%%%%%%%%%%%%%%%%%%%%%%%%%%%%%%%%%%%
 \begin{figure}
 {\begin{center}
 \includegraphics[width=80.0mm]{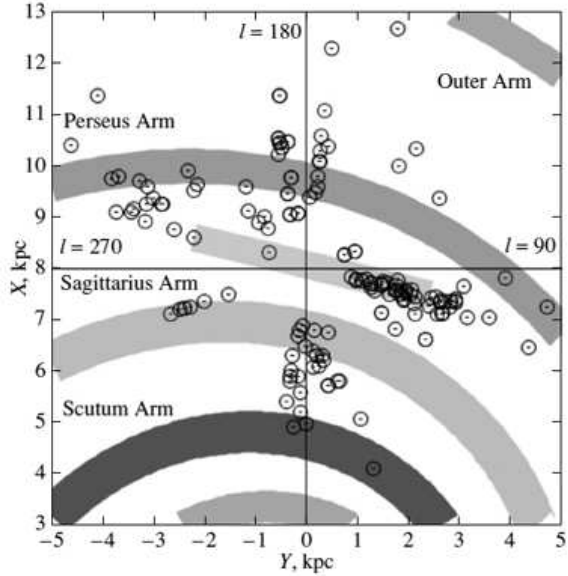} %%f-156-patri.eps}
 \caption{
Distribution of 156 O stars from the list by Patriarchi et al.
(2003) in projection onto the Galactic $XY$ plane; the Sun is at
the intersection of the dashed lines; the fragments of the
four-armed spiral pattern are indicated; the position of the local
spur, the Orion Arm, is marked. }
  \label{f-156-patri}
 \end{center} }
 \end{figure}
%%%%%%%%%%%%%%%%%%%%%%%%%%%%%%%%%%%%%%%%%%%%%%%%%%%%%%%%%%%%%%%%%%%
%%%%%%%%%%%%%%%%%%%%%%%%%%%%%%%%%%%%%%%%%%%%%%%%%%%%%%%%%%%%%%%%%%%
 \begin{figure}
 {\begin{center}
 \includegraphics[width=90.0mm]{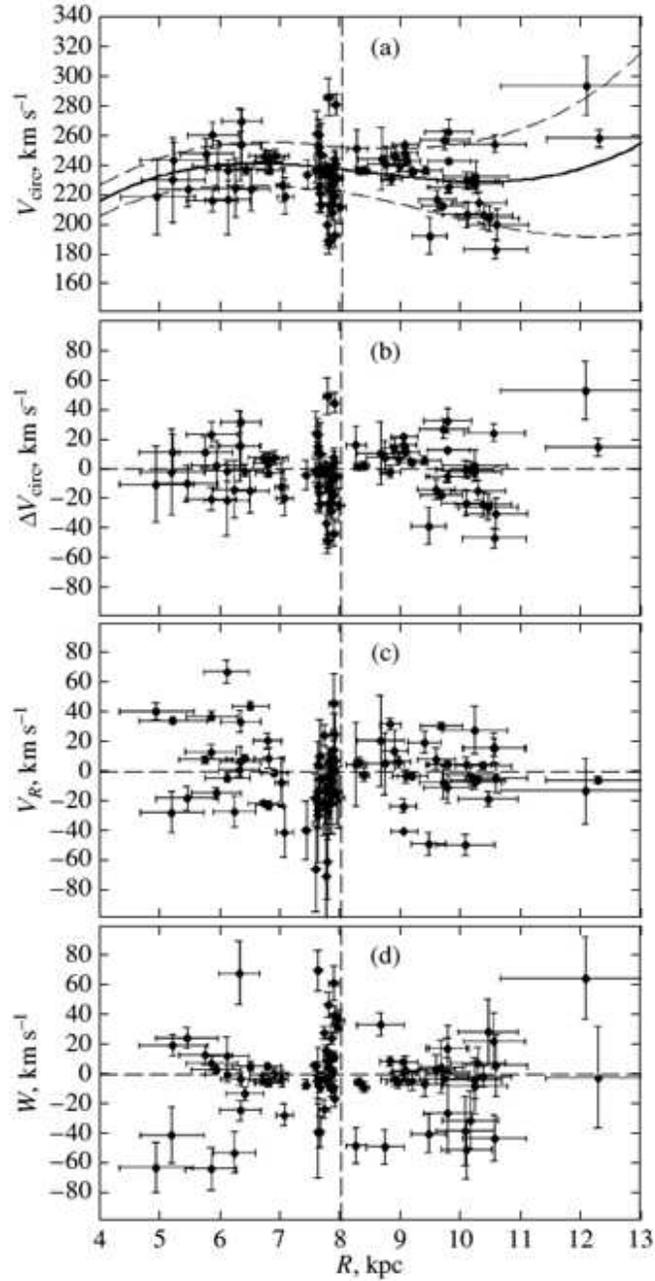}  %%f-55-patri.eps}
 \caption{
Galactic rotation curve constructed from the sample of 101 O stars
from the list by Patriarchi et al. (2003) with parameters (14)
with an indication of the boundaries of the 1$\sigma$ confidence
intervals (a); residual rotation velocities $\Delta V_{circ}$ (b),
radial velocities $V_R$ (c), and vertical velocities $W$ (d) of
stars versus $R;$ the vertical dashed line marks the Sun’s
position.}
  \label{f55-patri}
 \end{center} }
 \end{figure}
%%%%%%%%%%%%%%%%%%%%%%%%%%%%%%%%%%%%%%%%%%%%%%%%%%%%%%%%%%%%%%%%%%%

 \subsection*{The Distances Determined from Interstellar Ca II Lines}
The distances to OB stars determined by the spectroscopic method
from the broadening of interstellar Ca II, Na I, or K I absorption
lines are of indubitable interest. The method of determining such
distances is well known. However, the ``calcium'' distance scale
has been tied to the Hipparcos trigonometric parallaxes only
recently (Megier et al. 2005). The assumption about a uniform
distribution of ionized atoms in the Galactic plane underlies the
method. In some places (for example, toward the cluster Trumpler
16), the nonuniformities in the distribution of matter can be
significant, which can increase the dispersion of the distance
estimates by this method. On average, according to the estimates
by Megier et al. (2009), the accuracy of determining the
individual distances to OB stars is $\approx$15\%.

The first catalogue (Megier et al. 2009) contains 290 OB stars,
while Galazutdinov et al. (2015) determined the distances to 61
more OB stars by this method, with the line-if-sight velocities
having been measured for all of them. Since the two samples have
overlaps, the total number of stars in this distance scale is
about 340.

Previously (Bobylev and Bajkova 2011, 2013b), we studied a sample
of 258 Hipparcos O--B3 stars with distances from Megier et al.
(2009). About 20\% of the sample are either known runaway stars or
candidates for runaway stars, because they have large ($>40$~km
s$^{-1}$) residual space velocities. As a result, the range of
distances 0.8--3.5 kpc was represented by 102 stars. A slight, no
more than 20\%, reduction of the distance scale was shown to be
necessary by analyzing several kinematic parameters obtained from
this sample (Bobylev and Bajkova 2011). In this paper, we use the
entire sample, and not only the Hipparcos stars.

Figure 6 gives the distribution of 168 OB stars with distances in
the calcium distance scale in the Galactic $XY$ plane for two
distance scale factors. In the first case, this factor is equal to
one; in the second case, the distances to the stars were
multiplied by 0.8. Note that the new measurements by Galazutdinov
et al. (2015) were performed exclusively in two narrow sectors: in
directions $l\approx135^\circ$ and $l\approx190^\circ$ with
heliocentric distances of about 2.3 kpc. Two structures elongated
along the line of sight are clearly seen in Fig. 6 in these
directions.

\section*{RESULTS}
\subsection*{Spectroscopic Binary}
Using 120 spectroscopic binaries, we found the following kinematic
parameters from a least-squares solution of the system of
conditional equations (3)--(5):
 \begin{equation}
 \label{OMEGA-SB}
 \begin{array}{lll}
 (U_\odot,V_\odot,W_\odot)=(2.8,9.2,9.3)\pm(1.0,1.2,1.0)~\hbox{km s$^{-1}$},\\
      \Omega_0 =~29.3\pm0.8~\hbox{km s$^{-1}$ kpc$^{-1}$},\\
  \Omega^{'}_0 =-4.28\pm0.15~\hbox{km s$^{-1}$ kpc$^{-2}$},\\
 \Omega^{''}_0 =0.957\pm0.128~\hbox{km s$^{-1}$ kpc$^{-3}$}.
 \end{array}
 \end{equation}
In this solution, the error per unit weight is $\sigma_0=10.5$ km
s$^{-1}$. For the adopted $R_0=8.0\pm0.4$ kpc, the linear Galactic
rotation velocity $V_0=|R_0\Omega_0|$ is $234\pm14$ km s$^{-1}$,
while the Oort constants are $A=-17.1\pm0.6$~km s$^{-1}$
kpc$^{-1}$ and $B=12.2\pm1.0$~km s$^{-1}$ kpc$^{-1}$. The Galactic
rotation curve constructed with parameters (13) and the residual
tangential, $\Delta V_{circ},$ radial, $V_R,$ and vertical, $W,$
velocities of stars as a function of the distance R are given in
Fig. 2.

Why are the vertical velocities $W$ given in Fig.~2? The point is
that periodic vertical velocity oscillations have been detected
quite recently in Galactic masers with measured trigonometric
parallaxes (Bobylev and Bajkova 2015). Such oscillations with an
amplitude of about 4--6 km s$^{-1}$ can be associated with the
influence of the Galactic spiral density wave. Therefore,
confirming this phenomenon in the velocities of other samples of
stars with different distance scales is of great interest.

Based on a spectral analysis, we determined the parameters of the
spiral density wave using the sample of spectroscopic binary
stars. The results are reflected in Fig.~3, where the power
spectra of the radial, $V_R,$ residual tangential, $\Delta
V_{circ}$, and vertical, $W,$ velocities are given. Note that
there is a high significance of the signal, $p=0.999,$ only in the
power spectrum of the radial velocities. For the tangential
velocities, the significance of the signal at a wavelength
$\lambda\approx2.6$ kpc is $p=0.898.$ As can be seen on the middle
panel in Fig. 3, there are spurious signals in the range of short
wavelengths whose amplitudes exceed the amplitude of the signal of
interest to us. There are also spurious signals in the power
spectrum of the vertical velocities $W,$ while the signal at a
wavelength of about 1.5 kpc agrees poorly with the two preceding
graphs.

As a result, for the model of a four-armed spiral pattern $(m=4),$
we found the amplitudes of the velocity perturbations
 $f_R=9.5\pm1.5$ km s$^{-1}$ and
 $f_\theta=3.2\pm1.4$ km s$^{-1}$ in the radial
and tangential directions, respectively, the wavelength
  $\lambda_R=2.8\pm0.5$ kpc (then, $i_R=-13^\circ\pm4^\circ$) and
 $\lambda_\theta=2.6\pm0.4$ kpc (then, $i_\theta=-12^\circ\pm3^\circ$)
 at the Sun's phase in the spiral density wave
 $(\chi_\odot)_R=-95^\circ\pm15^\circ$, and
 $(\chi_\odot)_\theta=-93^\circ\pm12^\circ, $ respectively.
The corresponding waves are given on panels (b) and (c) in Fig. 2.
Their logarithmic character is clearly seen.

%%%%%%%%%%%%%%%%%%%%%%%%%%%%%%%%%%%%%%%%%%%%%%%%%%%%%%%%%%%%%%%%%%%
 \begin{figure}
 {\begin{center}
 \includegraphics[width=160.0mm]{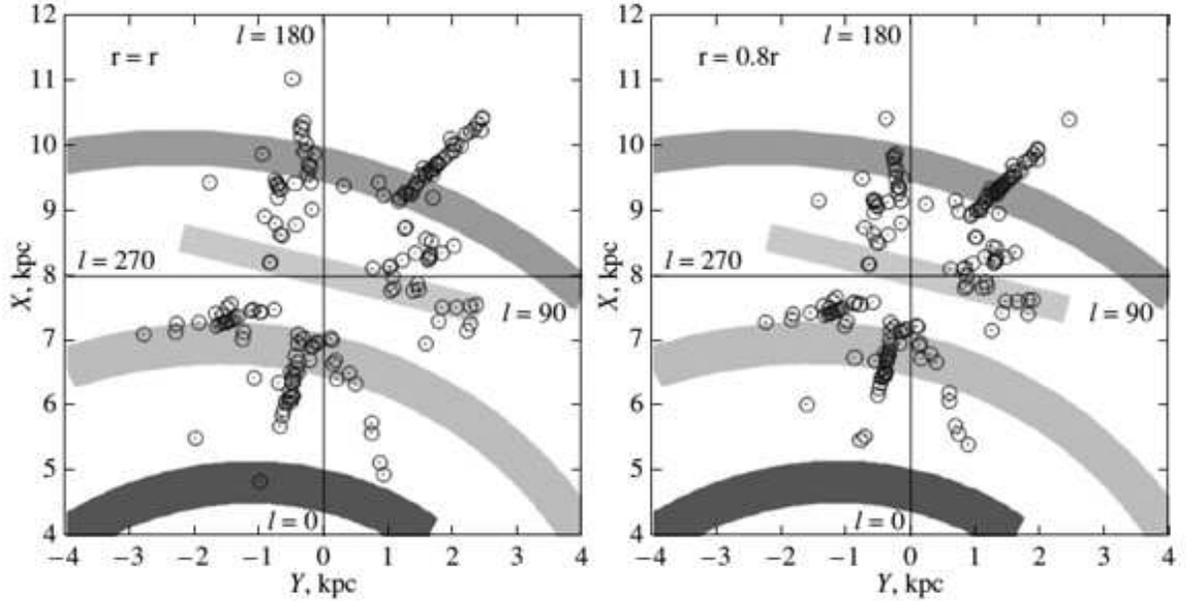}   %CaII-XY.eps}
 \caption{
Distribution of 168 OB stars with distances in the ``calcium''
distance scale in projection onto the Galactic $XY$ plane (a); the
distances to these stars were multiplied by factor of 0.8 (b); the
fragments of the four-armed spiral pattern and the position of the
Orion Arm are indicated. }
  \label{CaII-XY}
 \end{center} }
 \end{figure}
%%%%%%%%%%%%%%%%%%%%%%%%%%%%%%%%%%%%%%%%%%%%%%%%%%%%%%%%%%%%%%%%%%%
%%%%%%%%%%%%%%%%%%%%%%%%%%%%%%%%%%%%%%%%%%%%%%%%%%%%%%%%%%%%%%%%%%%
 \begin{figure}
 {\begin{center}
 \includegraphics[width=160.0mm]{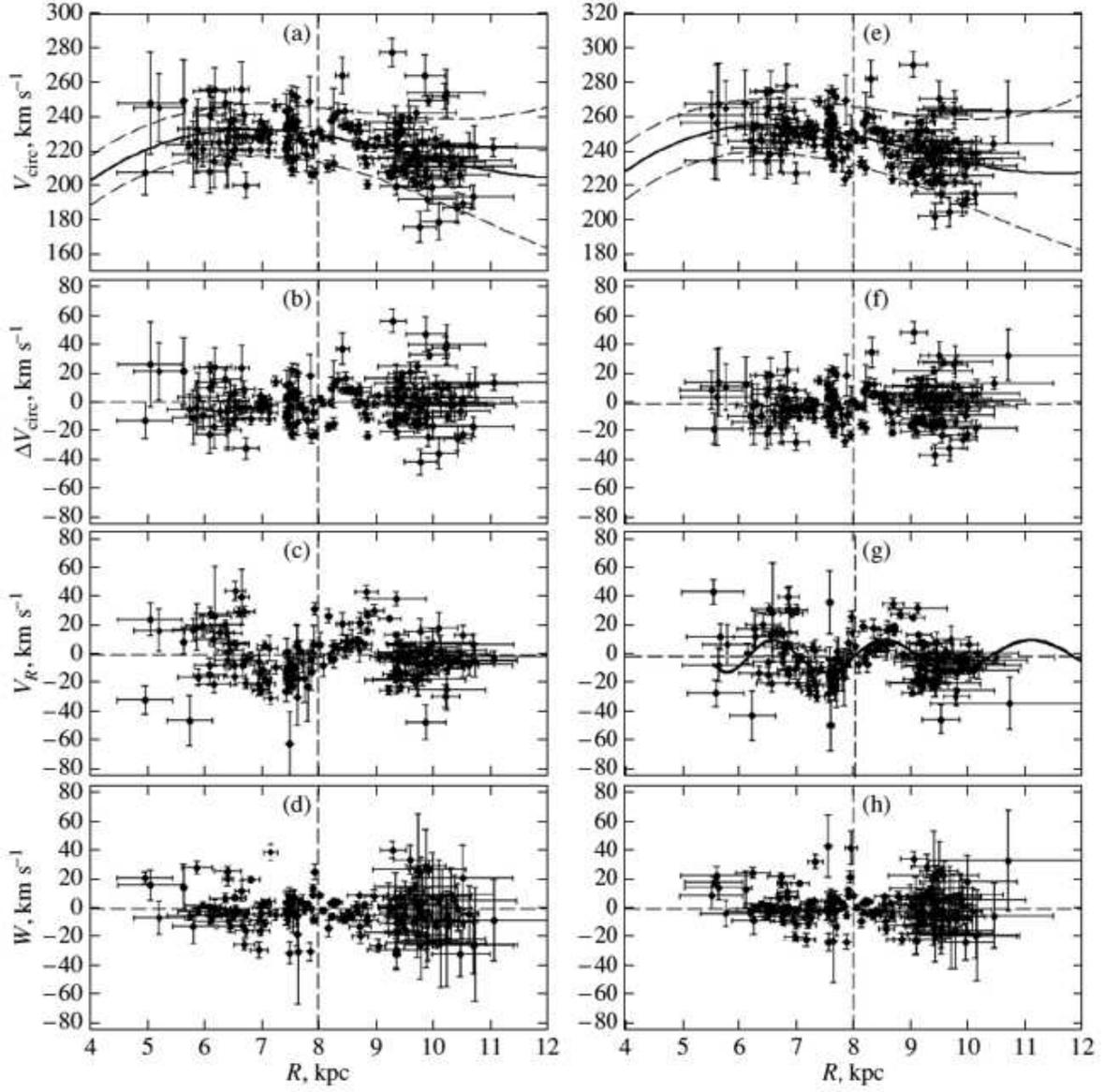}  %%CaII-rot-NEW.eps}
 \caption{
Galactic rotation curve constructed from 168 OB stars with
distances in the original ``calcium'' scale (a); residual rotation
velocities $\Delta V_{circ}$ (b), radial velocities $V_R$ (c), and
vertical velocities $W$ (d) of stars versus $R;$ the same
dependences are given on panels (e)--(g) for the stars whose
distances were reduced by 20\%.}
  \label{CaII-rotation}
 \end{center} }
 \end{figure}
%%%%%%%%%%%%%%%%%%%%%%%%%%%%%%%%%%%%%%%%%%%%%%%%%%%%%%%%%%%%%%%%%%%
%%%%%%%%%%%%%%%%%%%%%%%%%%%%%%%%%%%%%%%%%%%%%%%%%%%%%%%%%%%%%%%%%%%
 \begin{figure}
 {\begin{center}
 \includegraphics[width=60.0mm]{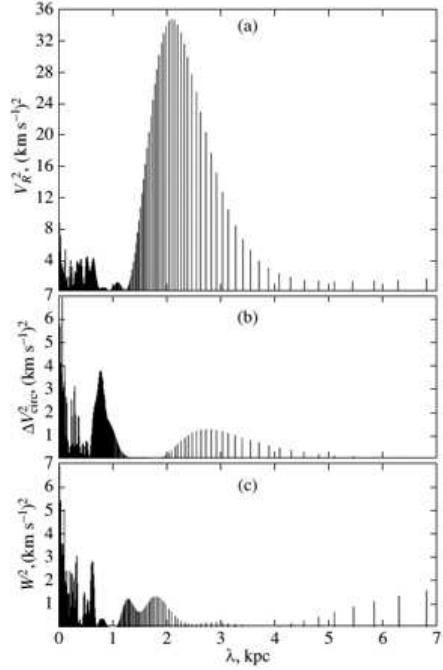}  %%Spectr-CaII.eps}
 \caption{
Power spectra of the radial, $V_R,$ residual tangential, $\Delta
V_{circ},$ and vertical, $W,$ velocities for the sample of OB3
stars with the calcium distance scale. }
  \label{Sp-CaII}
 \end{center} }
 \end{figure}
%%%%%%%%%%%%%%%%%%%%%%%%%%%%%%%%%%%%%%%%%%%%%%%%%%%%%%%%%%%%%%%%%%%

\subsection*{Stars from Patriarchi et al. (2003)}
Using 156 O stars from the list by Patriarchi et al. (2003), we
found the following parameters:
 \begin{equation}
 \label{OMEGA-156}
 \begin{array}{lll}
 (U_\odot,V_\odot,W_\odot)=(5.8,16.9,8.6)\pm(2.1,2.6,1.8)~\hbox{km s$^{-1}$},\\
      \Omega_0 =~29.6\pm1.2~\hbox{km s$^{-1}$ kpc$^{-1}$},\\
  \Omega^{'}_0 =-4.27\pm0.27~\hbox{km s$^{-1}$ kpc$^{-2}$},\\
 \Omega^{''}_0 =0.910\pm0.143~\hbox{km s$^{-1}$ kpc$^{-3}$}.
 \end{array}
 \end{equation}
In this solution, the error per unit weight is $\sigma_0=21.9$ km
s$^{-1}$, and the linear Galactic rotation velocity is
$V_0=237\pm15$ km s$^{-1}$ for $R_0=8.0\pm0.4$~kpc.

There are quite a few stars without line-of-sight velocities in
this list. Equations (3)--(5) can be solved with limited
information, which was done when obtaining solution (14). However,
complete information is needed to calculate the velocities $\Delta
V_{circ}, V_R,$ and $W,$ i.e., only 101 stars can be used.

The Galactic rotation curve constructed with parameters (14) and
the residual tangential, $\Delta V_{circ},$ radial, $V_R,$ and
vertical, $W,$ velocities of stars as a function of the distances
$R$ are given in Fig. 5. It can be seen from this figure that a
smooth Galactic rotation curve is determined quite well. Since the
velocities $\Delta V_{circ},$ $V_R,$ and $W$ are fairly irregular
in pattern, we did not perform a spectral analysis of these
velocities.

The error per unit weight $\sigma_0$ characterizes the dispersion
of the residuals in the least-squares solution of the system of
equations (3)--(5). Its value is usually close to the value of the
residual velocity dispersions for the sample stars averaged over
all directions (the ``cosmic'' velocity dispersion). The value of
$\sigma_0=21.9$ km s$^{-1}$ found in solution (14) is twice
$\sigma_0\approx10$ km s$^{-1}$ expected for OB stars. This may
imply that the errors in the distances and, consequently, the
tangential velocities of these stars are too great. Panel (d) in
Fig. 5, on which large vertical velocities $W$ are seen for a
considerable number of stars, attracts our attention.

We produced a new sample of stars with space velocities for two
constraints: $|W|<40$ km s$^{-1}$ and $|z|<100$ pc. Applying these
constraints allows the number of possible runaway stars to be
reduced significantly. 61 stars satisfy these criteria. Using
them, we found the following parameters:
 \begin{equation}
 \label{OMEGA-61-Patri}
 \begin{array}{lll}
 (U_\odot,V_\odot,W_\odot)=(10.1,14.6,5.7)\pm(2.1,2.3,1.9)~\hbox{km s$^{-1}$},\\
      \Omega_0 =~31.7\pm1.5~\hbox{km s$^{-1}$ kpc$^{-1}$},\\
  \Omega^{'}_0 =-4.53\pm0.29~\hbox{km s$^{-1}$ kpc$^{-2}$},\\
 \Omega^{''}_0 =1.124\pm0.247~\hbox{km s$^{-1}$ kpc$^{-3}$},
 \end{array}
 \end{equation}
where the error per unit weight is $\sigma_0=13.7$~km s$^{-1}$,
which is considerably smaller than that in solution (14).

\subsection*{The ``Calcium'' Distance Scale}
Based on 168 stars, we obtained the following solution for the
adopted $R_0=8.0\pm0.4$~kpc:
 \begin{equation}
 \label{OMEGA-r-1}
 \begin{array}{lll}
 (U_\odot,V_\odot,W_\odot)=\\
 (10.2,9.3,9.5)\pm(1.1,1.6,1.1)~\hbox{km s$^{-1}$},\\
      \Omega_0 =~28.7\pm0.9~\hbox{km s$^{-1}$ kpc$^{-1}$},\\
  \Omega^{'}_0 =-4.21\pm0.16~\hbox{km s$^{-1}$ kpc$^{-2}$},\\
 \Omega^{''}_0 =0.652\pm0.149~\hbox{km s$^{-1}$ kpc$^{-3}$},
 \end{array}
 \end{equation}
$\sigma_0=13.6$ km s$^{-1}$ and $V_0=230\pm15$ km s$^{-1}$. Based
on the same stars, we also obtained the solution for the distance
scale factor 0.8:
 \begin{equation}
 \label{OMEGA-r-0.8}
 \begin{array}{lll}
 (U_\odot,V_\odot,W_\odot)=\\
 (10.4,9.7,7.0)\pm(1.0,1.4,0.9)~\hbox{km s$^{-1}$},\\
      \Omega_0 =~31.3\pm0.9~\hbox{km s$^{-1}$ kpc$^{-1}$},\\
  \Omega^{'}_0 =-4.78\pm0.16~\hbox{km s$^{-1}$ kpc$^{-2}$},\\
 \Omega^{''}_0 =0.864\pm0.176~\hbox{km s$^{-1}$ kpc$^{-3}$},
 \end{array}
 \end{equation}
$\sigma_0=11.6$~km s$^{-1}$ и $V_0=250\pm15$~km s$^{-1}$. Figure 7
gives the Galactic rotation curve and the velocities $\Delta
V_{circ},$ $V_R,$ and $W$ as a function of R constructed from OB
stars with distances in the original calcium scale and the scale
reduced by 20\%.

Based on a spectral analysis, we determined the parameters of the
spiral density wave from the sample of spectroscopic binary stars
with the reduced distance scale. The results are reflected in Fig.
8, where the power spectra of the radial, $V_R,$ residual
tangential, $\Delta V_{circ},$ and vertical, $W,$ velocities are
given. Note that there is a high significance of the signal,
$p=0.999,$ only in the power spectrum of the radial velocities. It
can be clearly seen on panels (b) and (c) that there is no
statistically significant signal with a wavelength of about
2.5~kpc typical of the spiral density wave for the tangential and
vertical velocities.

As a result, from our analysis of the radial velocities we found
the velocity perturbation amplitude $f_R=11.8\pm1.3$ km s$^{-1}$
and the wavelength $\lambda=2.1\pm0.4$ kpc
($i=-9.5^\circ\pm1.7^\circ$) at the Sun's phase in the spiral
density wave $\chi_\odot=-86^\circ\pm7^\circ$. This wave is given
on panel (g) in Fig. 7.

\section*{DISCUSSION}
First of all, it should be noted that the peculiar solar velocity
components, the angular velocity of Galactic rotation, and its two
derivatives, are determined quite well from all three samples of
stars. They are determined with the smallest errors from the
sample of spectroscopic binary stars and the sample of stars with
the calcium distance scale.

For two samples, we also determined such kinematic parameters
previously, but only using a smaller number of stars. The results
of solution (13) should be compared with the results of our
analysis for 58 distant spectroscopic binary stars outside the
circle with a radius of 0.6 kpc (Bobylev and Bajkova 2013a):
$(U_\odot,V_\odot)= (2.2,8.4)\pm(0.9,1.1)$~km s$^{-1}$ and
 $\Omega_0      = 31.9 \pm1.1$~km s$^{-1}$ kpc$^{-1}$,
 $\Omega^{'}_0  = -4.30\pm0.16$~km s$^{-1}$ kpc$^{-2}$,
 $\Omega^{''}_0 =  1.05\pm0.35$~km s$^{-1}$ kpc$^{-3}$,
where the error per unit weight is $\sigma_0=9.9$~km s$^{-1}$ and
$V_0=255\pm16$~km s$^{-1}$.

Based on a sample of 102 OB3 stars with the calcium distance scale
(and the scale factor 0.8), previously (Bobylev and Bajkova 2011)
we found the following parameters:
 $(U_\odot,V_\odot,W_\odot)=(8.9,10.3,6.8)\pm(0.6,1.0,0.4)$~km s$^{-1}$,
 $\Omega_0 = 31.5\pm0.9$~km s$^{-1}$ kpc$^{-1}$,
 $\Omega^{'}_0 = -4.49\pm0.12$~km s$^{-1}$ kpc$^{-2}$,
 $\Omega^{''}_0 = 1.05\pm0.38$~km s$^{-1}$ kpc$^{-3}$,
 the error per unit
weight was $\sigma_0=9.6$~km s$^{-1}$, and the circular rotation
velocity was $V_0=252\pm14$ km s$^{-1}$ ($R_0=8$ kpc). We can see
good agreement with solution (17). The only difference is that the
error in the second derivative of the angular velocity
$\Omega^{''}_0$ decreased considerably with increasing number of
stars.

The distances for the O stars from the list by Patriarchi et al.
(2003) were determined not quite reliably. Note that these authors
took the absolute magnitudes of the stars $M_v$ from the rather
old paper by Panagia (1973) and their spectral types from the
catalogue by Garmany et al. (1982). The correct spectral
classification is of great importance in determining the
spectroscopic distances of stars. A telling example is the star HD
160641, for which the spectral type O9.5I is specified in the
catalogue by Patriarchi et al. (2003), and the distance $r=9$ kpc
was calculated. According to the estimates by a number of authors,
this is a helium star, a sdOc9.5II-III:He40 dwarf (Drilling et al.
2013) with a mass of $\sim$1$M_\odot$. In this case, the distance
to the star is $r=3.3\pm0.8$~kpc (Lynas-Gray et al. 1987).

Why does the sample of spectroscopic binaries look better? After
all, the distances to the stars were determined either by the
photometric method or by the spectroscopic one. In our view, the
fact that each star is ``piece goods'' when the spectroscopic
orbits are determined plays a role here. In this case, the authors
have a detailed idea of the binary characteristics, in particular,
of the spectra of the binary components; therefore, the
inaccuracies in the spectral classification are minimal.

At present, there is a sample of approximately 100 maser sources
whose trigonometric parallaxes have been measured by the VLBI
method with a very high accuracy, with a mean error of $\pm20$
$\mu$as and, for some of them, with a record error of
$\pm5~\mu$as. From an analysis of these masers, Reid et al. (2014)
found the velocity of the Sun $V_0=240\pm8$ km s$^{-1}$
($R_0=8.34\pm0.16$ kpc). Previously, based on a smaller number of
masers, Honma et al. (2012) obtained an estimate of $V_0=238\pm14$
km s$^{-1}$ ($R_0=8.05\pm0.45$ kpc). It can be seen that the
values of this velocity found in this paper are in good agreement
with the best present day estimates.

Based on a sample of masers, previously (Bobylev and Bajkova 2015)
we found the following parameters of the spiral density wave for
$m=4:$  $f_\theta=6.0\pm2.6$~km s$^{-1}$,
        $f_R=7.2\pm2.2$~km s$^{-1}$,
   $\lambda_\theta=3.2\pm0.5$~kpc,
   $\lambda_R=     3.0\pm0.6$~kpc,
   $(\chi_\odot)_\theta= -79^\circ\pm14^\circ$ and
   $(\chi_\odot)_R=-199^\circ\pm16^\circ$.
 The values found in this paper from the
sample of spectroscopic binary stars are fairly close to them. The
values found from the radial velocities of stars with the calcium
distance scale are characterized by a larger amplitude and a
smaller $(\chi_\odot)_R.$ However, this just confirms the results
of our previous analysis (Bobylev and Bajkova 2011).

%%%%%%%%%%%%%%%%%%%%%%%%%%%%%%%%%%%%%%%%%%%%%%%%%%%%%%%%%%%%%%%%%%%
 \begin{figure}
 {\begin{center}
 \includegraphics[width=80.0mm]{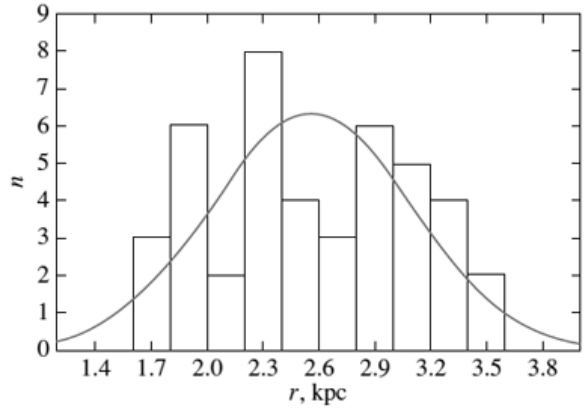}  %%SigDist-CaII.eps}
 \caption{
Distribution of 43 stars with the calcium distance scale on the
line of sight in the direction $l=135^\circ;$ a Gaussian with a
mean of 2.56 pc and a dispersion of 0.54 pc was fitted.}
  \label{SigDist-CaII}
 \end{center} }
 \end{figure}
%%%%%%%%%%%%%%%%%%%%%%%%%%%%%%%%%%%%%%%%%%%%%%%%%%%%%%%%%%%%%%%%%%%

Given the rather significant relative errors in the distances for
many of the stars being used, the question about the systematic
bias in the mean distances of the remaining stars arises.
Obviously, there are relatively more objects with overestimated
distances among the rejected stars farther than 4 kpc and, on the
contrary, with underestimated distances among the stars nearer
than 600 pc. Such a bias in the distances is related to the
Lutz–Kelker effect (Lutz and Kelker 1973; Stepanishchev and
Bobylev 2013). For the classical case where a uniform distribution
of stars is adopted, we can use a formula of the form (Lutz and
Kelker 1973)
 \begin{equation}
 \label{Lutz Kelker}
 F(z)={1\over z^4} \exp {\biggl[-{(z-1)^2 \over 2\sigma_p^2} \biggr]},
 \end{equation}
where $z=\pi/\pi_0$ is the ratio of the true parallax $\pi$ to the
observed parallax $\pi_0,$  $\sigma_p=\sigma_0/\pi_0$ is the
relative error of the observed parallax, and $F(z)$ is the
probability density.

We took the mean error in the distances for the spectroscopic
binaries and the stars with the calcium distance scale to be 15\%.
The mean correction for the Lutz--Kelker effect for these stars is
then 100 pc (the measured distances should be increased). For the
stars from the list by Patriarchi et al. (2003), we took the error
in the distance for each star to be 30\%. As a result of our
simulations, we found that the total bias for these stars due to
the Lutz--Kelker effect could be 240 pc.

The chains of stars elongated along the line of sight, which can
be seen in Fig.~4 and particularly clearly in Fig.~6, deserve
special attention. The most plausible explanation is that there
are large relative errors in the distances to the objects that
actually belong to a single compact grouping (a cluster, an
association, or a stellar complex). Based on such objects, we
obtained an independent estimate of the relative error in the
distance using stars with the calcium distance scale as an
example. The result is reflected in Fig.~9, where the distribution
of 43 OB stars with the calcium distance scale lying on the line
of sight in the direction $l=135^\circ$ is given. Based on the
parameters of the fitted Gaussian (a mean of 2.56 pc and a
dispersion of 0.54 pc), we find the relative error in the distance
to be 21\%. The mean value calculated from the individual distance
errors for these stars is 0.42; the relative error in the distance
is then 16\%.

\section*{CONCLUSIONS}
We considered O- and B-type stars located in a wide solar
neighborhood, with distances from 0.6 to 4 kpc. We produced three
samples of stars for which the distances, line-of-sight
velocities, and proper motions collected from published data were
available. The first sample includes 120 massive (with masses $>10
M_\odot$) spectroscopic binaries. O stars for which Patriarchi et
al. (2003) estimated the spectroscopic distances using infrared
photometric data from the 2MASS catalogue constitute the second
sample. Single stars constitute the overwhelming majority in this
sample. The third sample consists of 168 OB3 stars whose distances
were determined from interstellar calcium lines. As in the
previous case, mostly single stars enter into this sample.

Such kinematic parameters as the angular velocity of Galactic
rotation at the solar distance $\Omega'_0,$ its two derivatives
$\Omega'_0,$ and $\Omega''_0$ and the peculiar solar velocity
components $(U,V,W)_\odot$ were shown to be well determined from
all three samples of stars. They are determined with the smallest
errors from the sample of spectroscopic binaries and the sample of
stars with the calcium distance scale.

The fine structure of the velocity field associated with the
influence of the Galactic spiral density wave clearly manifests
itself in the radial velocities of the spectroscopic binaries and
in the sample of stars with the calcium distance scale. No
small-amplitude periodic oscillations in the vertical velocities
with a wavelength of about 2.5 kpc manifest themselves in any of
the stellar samples considered in this paper. The available
accuracies of the distances are probably insufficient for a
reliable detection of such oscillations.

The linear rotation velocity of the Galaxy $V_0$ was found from
the sample of spectroscopic binary stars to be $234\pm14$~km
s$^{-1}$ (for the adopted $R_0=8.0\pm0.4$ kpc). Based on a
spectral analysis of the radial and tangential velocities for
these stars, we determined the parameters of the spiral density
wave. For the model of a four-armed spiral pattern $(m=4),$ we
found the following: the perturbation velocity amplitudes
 $f_R =9.5\pm1.5$ km s$^{-1}$ and
 $f_\theta=3.2\pm1.4$ km s$^{-1}$ in the
radial and tangential directions, respectively; the wavelength
 $\lambda_R=2.8\pm0.5$ kpc (then, $i_R=-13^\circ\pm4^\circ$) at the Sun's phase in
the spiral density wave $(\chi_\odot)_R=-95^\circ\pm15^\circ$ and
 $\lambda_\theta=2.6\pm0.4$ kpc (then, $i_\theta=-12^\circ\pm3^\circ$)
 at the Sun's phase $(\chi_\odot)_\theta=-93^\circ\pm12^\circ.$

The velocity $V_0=250\pm15$~km s$^{-1}$ (for $R_0=8.0\pm0.4$ kpc
and a distance scale factor of 0.8) was found from the sample of
OB3 stars with the calcium distance scale. The parameters of the
spiral density wave were determined only by analyzing the radial
velocities. The perturbation velocity amplitude is
 $f_R=11.8\pm1.3$~km s$^{-1}$ and the wavelength is
 $\lambda_R=2.1\pm0.3$~kpc ($m=4,$ $i_R=-9.5^\circ\pm1.7^\circ$)
at the Sun's phase in the spiral density wave
  $(\chi_\odot)_R=-86^\circ\pm7^\circ$.

\subsection*{ACKNOWLEDGMENTS} We are grateful to the referees for their useful
remarks that contributed to an improvement of the paper. This work
was supported by the ``Transient and Explosive Processes in
Astrophysics'' Program P--41 of the Presidium of the Russian
Academy of Sciences. The SIMBAD electronic astronomical database
was widely used in our work.

 \newpage
 \bigskip{REFERENCES}
 \bigskip
 {\small

 1. L.D. Anderson, T.M. Bania, D.S. Balser, and R.T. Rood, Astrophys. J. 754, 62 (2012).

2. J.I. Arias, R.H. Barb\'a, R.C. Gamen, N.I. Morrell, J.
Maiz-Apell\'aniz, E.J. Alfaro, A. Sota, N.R. Walborn, and C. M.
Bidin, Astrophys. J. Lett. 710, L30 (2010).

 3. V.S. Avedisova and G.I. Kondratenko, Nauchn. Inform. Astron. Sov. AN SSSR 56, 59 (1984).

 4. V.S. Avedisova, Astron. Rep. 49, 435 (2005).

 5. A.T. Bajkova and V.V. Bobylev, Astron. Lett. 38, 549 (2012).

6. V. Baki\c{s}, H. Hensberge, S. Bilir, H. Baki\c{s}, F. Yilmaz,
E. Kiran, O. Demircan, M. Zejda, and Z. Mikula\v{s}ek, Astron. J.
147, 149 (2014).

 7. V.V. Bobylev and A.T. Bajkova, Astron. Lett. 37, 526 (2011).

 8. V.V. Bobylev and A.T. Bajkova, Astron. Lett. 39, 532 (2013a).

 9. V.V. Bobylev and A.T. Bajkova, Astron. Nachr. 334, 768 (2013b).

 10. V.V. Bobylev and A.T. Bajkova, Mon. Not. R. Astron. Soc. 437, 1549 (2014a).

 11. V.V. Bobylev and A.T. Bajkova, Astron. Lett. 40, 783 (2014).

 12. V.V. Bobylev and A.T. Bajkova, Mon. Not. R. Astron. Soc. 447, L50 (2015).

 13. C. T. Bolton and G. L. Rogers, Astrophys. J. 222, 234 (1978).

 14. D. V. Bowen, E.B. Jenkins, T.M. Tripp, K.R. Sembach, B. D.
Savage, H.W. Moos, W.R. Oegerle, S.D. Friedman, et al., Astrophys.
J. Suppl. Ser. 176, 59 (2008).

15. G. Capilla and J. Fabregat, Astron. Astrophys. 394, 479
(2002).

16. A. Carmona, M.E. van den Ancker, M. Audard, Th. Henning, J.
Setiawan, and J. Rodmann, Astron. Astrophys. 517, 67 (2010).

17. J. Casares, I. Negueruela, M. Rib\'o, I. Ribas, J.M. Paredes,
A. Herrero, and S. Sim\'on-Diaz, Nature 505, 378 (2014).

 18. \"O. \c{C}hakirli, C. Ibanoglu, and E. Sipahi, Mon. Not. R. Astron. Soc. 442, 1560 (2014a).

 19.~\"O. \c{C}hakirli, C. Ibanoglu, E. Sipahi, A. Frasca, and G. Catanzaro, arXiv:1406.0499 (2014b).

 20.~\"O. \c{C}hakirli, New Astron.  35, 71 (2015).

21. A. Coleiro and S. Chaty, Astrophys. J. 764, 185 (2013).

22. F. Comer\'on and A. Pasquali, Astron. Astrophys. 543, 101
(2012).

23. S.M. Dougherty, A.J. Beasley, M.J. Claussen, B.A. Zauderer,
and N.J. Bolingbroke, Astrophys. J. 623, 447 (2005).

24. J.S. Drilling, C.S. Jeffery, U. Heber, S. Moehler, and R.
Napiwotzki, Astron. Astrophys. 551, 31 (2013).

25. S. Dzib and L.F. Rodriguez, Rev. Mex. Astron. Astrofis. 45, 3
(2009).

26. Yu.N. Efremov, Astron.Rep. 55, 108 (2011).

27. Z. Eker, S. Bilir, F.~Soydugan, E.Y. G\"ok\c{c}e, M. Soydugan,
 M. T\"uys\"uz, T.~\c{S}eny\"uz, and O. Demircan, Publ. Astron. Soc.
Austral. 31, 23 (2014).

28. G.A. Galazutdinov, A. Strobel, F.A. Musaev, A. Bondar, and J.
Krelowski, Publ. Astron. Soc. Pacif. 127, 126 (2015).

29. C.D. Garmany, P.S. Conti, and C. Chiosi, Astrophys. J. 263,
777 (1982).

30. G.A. Gontcharov, Astron. Lett. 32, 795 (2006).

31. S.B. Gudennavar, S.G. Bubbly, K. Preethi, and J. Murthy,
Astrophys. J. Suppl. Ser. 199, 8 (2012).

 32. E.F. Guinan, P. Mayer, P. Harmanec, H. Bo\v{z}i\'c, M. Bro\v{z}, J. Nemravov\'a,
 S. Engle, et al., Astron. Astrophys. 546, 123 (2012).

33. G.Hill and W.A. Fisher,Astron. Astrophys. 171, 123 (1987).

34. E. Hog, C. Fabricius, V.V. Makarov, S. Urban, T. Corbin, G.
Wycoff, U. Bastian, P. Schwekendiek, and A. Wicenec, Astron.
Astrophys. 355, L 27 (2000).

 35. M. Honma, T. Nagayama, K. Ando, T. Bushimata, Y.K. Choi, T. Handa, et al.,
 Publ. Astron. Soc. Jpn. 64, 136 (2012).

36. L.G. Hou and J.L. Han, Astron. Astrophys. 569, 21 (2014).

37. J.B. Hutchings and R.O. Redman, Mon. Not. R. Astron. Soc. 163,
219 (1971).

38. C. Ibanoglu, \'O. \c{C}akirli, and E. Sipahi, Mon. Not. R.
Astron. Soc. 436, 750 (2013).

39. N. Kaltcheva and M. Scorcio, Astron. Astrophys. 514, 59
(2010).

 40. D.C. Kiminki, H.A. Kobulnicky, K. Kinemuchi, J.S. Irwin, et al.,
 Astrophys. J. 664, 1102 (2007).

41. D.C. Kiminki, H.A. Kobulnicky, I. Gilbert, S. Bird, and G.
Chunev, Astron. J. 137, 4608 (2009).

42. D.C. Kiminki, H.A. Kobulnicky, I. Ewing, M.M.B. Kiminki, M.
Lundquist, M. Alexander, C. Vargas-Alvarez, H. Choi, and C.B.
Henderson, Astrophys. J. 747, 41 (2012).

43. F. van Leeuwen, Astron. Astrophys. 474, 653 (2007).

44. C.C. Lin and F.H. Shu, Astrophys. J. 140, 646 (1964).

45. A. Lobel, J. Groh, C. Martayan, Y. Fr\'emat, K.T. Dozinel, G.
Raskin, et al., Astron. Astrophys. 559, 16 (2013).

46. J. Lorenzo, I.Negueruela, A.K.F. val Baker, M. Garcia, S.
Sim\'on-Diaz, P. Pastor, and M. M\'endez Majuelos, Astron.
Astrophys. 572, 110 (2014).

47. T.E. Lutz and D.H. Kelker,Publ. Astron. Soc. Pacif. 85, 573
(1973).

 48. A.E. Lynas-Gray, D. Kilkenny, I. Skillen, and C. F.
Jeffery, Mon. Not. R. Astron. Soc. 227, 1073 (1987).

49. J. Maiz-Apell\'aniz, N.R. Walborn, N.I. Morrell, V.S. Niemela,
and E.P. Nelan, Astrophys. J. 660, 1480 (2007).

50. S.L. Malchenko, A.E. Tarasov, and K. Yakut, Odessa Astron.
Publ. 20, 120 (2007).

51. P. Mayer, H. Drechsel, and A. Irrgang, Astron. Astrophys. 565,
86 (2014).

52. M.V. McSwain, Astrophys. J. 595, 1124 (2003).

53. A. Megier, A. Strobel, A. Bondar, F.A. Musaev, I. Han, J.
Krelowski, and G.A. Galazutdinov, Astrophys. J. 634, 451 (2005).

54. A. Megier, A. Strobel, G.A. Galazutdinov, and J. Krelowski,
Astron. Astrophys. 507, 833 (2009).

 55. J.C.A. Miller-Jones, Publ. Astron. Soc. Austral. 31, 16 (2014).

 56. A.E.J. Moffat, S.V. Marchenko, W. Seggewiss, K.A. van der Hucht, H. Schrijver,
 B. Stenholm, et al., Astron. Astrophys. 331, 949 (1998).

57. A.P. Mois\'es, A. Damineli, E. Figueredo, R.D. Blum, P. Conti,
and C. L. Barbosa, Mon. Not. R. Astron. Soc. 411, 705 (2011).

58. A. Nasseri, R. Chini, P. Harmanec, P. Mayer, J.A. Nemravov\'a,
T. Dembsky, H. Lehmann, H. Sana, and J.-B. le Bouquin, Astron.
Astrophys. 568, 94 (2014).

59. N. Panagia, Astron. J. 78, 929 (1973).

60. P. Patriarchi, L. Morbidelli, and M. Perinotto, Astron.
Astrophys. 410, 905 (2003).

61. M.E. Popova and A.V. Loktin, Astron. Lett. 31, 663 (2005).

62. D.M. Popper, Publ. Astron. Soc. Pacif. 89, 315 (1977).

63. D. Pourbaix, A.A. Tokovinin, A.H. Batten, F.C. Fekel, W.I.
Hartkopf, H. Levato, N.I. Morell, G. Torres, and S. Udry, Astron.
Astrophys. 424, 727 (2004).

64. G. Rauw, H. Sana, I.I. Antokhin, N.I. Morrell, V.S. Niemela,
J.F.A. Colombo, E. Gosset, and J.-M. Vreux, Mon. Not. R. Astron.
Soc. 326, 1149 (2001).

65. M.J. Reid, K.M. Menten, A. Brunthaler, X.W. Zheng, T.M. Dame,
Y. Xu, et al., Astrophys. J. 783, 130 (2014).

66. A.C. Rovero and A.E. Ringuelet, Mon. Not. R. Astron. Soc. 266,
203 (1994).

67. H. Sana, E. Gosset, and C.J. Evans, Mon. Not. R. Astron. Soc.
400, 1479 (2009).

68. H. Sana, J.-B. Le Bouquin, L. Mahy, O. Absil, M. De Becker,
and E. Gosset, Astron. Astrophys. 553, 131 (2013).

69. M.F. Skrutskie, R.M. Cutri, R. Stiening, et al., Astron. J.
131, 1163 (2006).

70. N. Smith, R.D. Gehrz, O. Stahl, B. Balick, and A. Kaufer,
Astrophys. J. 578, 464 (2002).

71. A.S. Stepanishchev and V.V. Bobylev, Astron. Lett. 39, 185
(2013).

72. D.J. Stickland and C. Lloyd, Observatory 119, 16 (1999).

73. K.A. Stoyanov, R.K. Zamanov, G.Y. Latev, A.Y. Abedin, and N.A.
Tomov, Astron. Nachr. 335, 1060 (2014).

74. V. Straizys, Multicolor Stellar Photometry (Pachart, Tucson,
1992).

75. A. Tkachenko, P. Degroote, C. Aerts, K. Pavlovski, J.
Southworth, P.I. P\'apics, et al., Mon. Not. R. Astron. Soc. 438,
3093 (2014).

76. G. Torres, J. Andersen, and A. Gim\'enez, Astron. Astrophys.
Rev. 18, 67 (2010).

77. M. T\"uys\"uz, F. Soydugan, S. Bilir, E. Soydugan,
 T. Seny\"uz, and T. Yontan, New Astron. 28, 44 (2014).

78. J.P. Vall\'ee, Mon. Not. R. Astron. Soc. 442, 2993 (2014).

79. W. Wegner, Mon. Not. R. Astron. Soc. 374, 1549 (2007).

80. S.J. Williams, D.R. Gies, T.C. Hillwig, M.V. McSwain, and W.
Huang, in Proceedings of the Conference on Massive Stars: From
$\alpha$ to $\Omega$, June 10--14, 2013, Rhodes, Greece (2013).

81. B. Ya\c{s}arsoy and K. Yakut, Astrophys. J. 145, 9 (2013).

82. M.V. Zabolotskikh, A.S. Rastorguev, and A.K. Dambis, Astron.
Lett. 28, 454 (2002).

83. N. Zacharias, C.T. Finch, T.M. Girard, A. Henden, J.L.
Bartlett, D.G. Monet, and M.I. Zacharias, Catalogue No. I/322,
Strasbourg DataBase (2012).

84. The Hipparcos and Tycho Catalogues, ESA SP--1200 (1997).

 }

\end{document}